\documentclass{svproc}

\usepackage[utf8]{inputenc}
\usepackage{graphicx}
\usepackage{url}

\begin{document}
\mainmatter              
\title{Graph-based Retrieval for Claim Verification over Cross-Document Evidence}
\titlerunning{Cross-Document Claim Verification}  
%
\author{Misael Mongiov\`i 
\and Aldo Gangemi 
}
\authorrunning{Misael Mongiov\`i et al.} 
%
%
\institute{ISTC-CNR, Catania and Rome, Italy\\
\email{\{name\}.\{surname\}@istc.cnr.it}
}

\maketitle              

\begin{abstract}
Verifying the veracity of claims requires reasoning over a large knowledge base, often in the form of corpora of trustworthy sources. A common approach consists in retrieving short portions of relevant text from the reference documents and giving them as input to a natural language inference module that determines whether the claim can be inferred or contradicted from them. This approach, however, struggles when multiple pieces of evidence need to be collected and combined from different documents, since the single documents are often barely related to the target claim and hence they are left out by the retrieval module. 
We conjecture that a graph-based approach can be beneficial to identify fragmented evidence. We tested this hypothesis by building, over the whole corpus, a large graph that interconnects text portions by means of mentioned entities and exploiting such a graph for identifying candidate sets of evidence from multiple sources. Our experiments show that leveraging on a graph structure is beneficial in identifying a reasonably small portion of passages related to a claim.
\end{abstract}

\section{Introduction}

Claim verification~\cite{thorne_fever_2018} is the task of deciding whether a claim is supported or refuted (or neither of them) by a reference knowledge base. It is a fundamental task in automated fact checking~\cite{thorne_automated_2018}, with notable implications in the critical problem of contrasting disinformation, which has a significant worldwide impact~\cite{vosoughi_spread_2018}.     
Besides automatic fact checking, this task is relevant in every situation where the consistency of statements with respect to a knowledge base need to be verified. For instance, keeping a knowledge base updated when new information is available requires a consistency check of new data with previous knowledge. Integrating a supported claim would introduce redundancy, while a contradicted claim would make the knowledge base inconsistent. This is particularly important in robotics, in a scenario where a robot continuously updates its knowledge through conversation with humans. New statements that are supported by its internal knowledge might increase the confidence on encoded facts, while statements that are refuted might indicate misunderstanding, unreliability of the interlocutor or the presence of false facts in the knowledge base.    

Although modern language models are reasonably effective in deciding whether a short text (premise) infers or contradicts a claim (hypothesis), their performances fall when the size of the premise increases. Therefore, effective retrieval of related sentences from a reference corpus is crucial for accurate verification. Recent approaches on claim verification are indeed based on a retrieve-and-verify paradigm where given a claim, text passages related to the claim are first retrieved, then a classifier is employed to decide whether the retrieved text supports, refutes or is neutral about the claim.
Thorne et al.~\cite{thorne_fever_2018} describe a general framework for verifying claims over a large corpus of text documents consisting in three steps. The \emph{document retrieval} step extracts from the corpus a set of documents related to the claim, which are likely to contain the evidence for or against the claim. Then a \emph{sentence selection} step performs a refinement by identifying in the retrieved documents a restricted set of sentences related to the claim, which we call the evidence. The last step, \emph{claim verification}, decides whether the claim or its negation can be inferred by the selected set of sentences and classifies the claim into one of the labels \emph{supported}, \emph{refuted} or \emph{not enough info}. Recently proposed claim verification methods are mostly based on this framework~\cite{zhou_gear_2019,zhong_reasoning_2020,yoneda_ucl_2018,soleimani_bert_2020,nie_combining_2019,ma_sentence-level_2019,liu_fine-grained_2021-1,hanselowski_ukp-athene_2018}.

Although the described framework achieves significant accuracy on average, it struggles when multiple pieces of evidence, some of them little related to the claim, are distributed across different documents. As a toy example, the claim ``The Beatles were formed in England'' can be verified by the following two sentences: ``The Beatles were formed in Liverpool'' and ``Liverpool is a city and metropolitan borough in Merseyside, England''. If we consider Wikipedia as the reference corpus, such sentences appear in separate documents, i.e. the page about ``The Beatles'' and the page about ``Liverpool''. While ``The Beatles'' is clearly related to the claim, the relevance of ``Liverpool'' is negligible since it is not mentioned in the claim. Without other knowledge, its relevance would be no higher then any other city in England. Of course the Liverpool page can be identified by implicit or explicit background knowledge, but in general such knowledge is not always available, and even when it is, it might be considered unreliable.
Some recent work (e.g.~\cite{lewis_retrieval_2021}) focus explicitly on the problem of retrieving documents related to a sentence (a claim or a question). Although they consider the distribution of related passages across documents, they do not give a specific solution and hence they still struggle when the evidence is fragmented into several document, each of them loosely related to the claim.

We propose a graph-based approach to retrieve relevant evidence for claim verification that translates the problem into a network search problem. The key idea is that the reference corpus can be summarized by a network that connects mentioned \emph{entities} (named entities or, in general, entities that can be uniquely identified) that are referred contextually, whose exploration can help identifying relevant concepts and corresponding text portions. In the example above, the path that connects the mentions to ``The Beatles", ``Liverpool"  and ``England" in the reference corpus outlines the evidence for the claim.

In the remainder we first describe our graph-based approach for retrieving evidence across documents (Section~\ref{sec:method}), then we report the results of our experimental analysis (Section~\ref{sec:experiments}). We discuss related work (Section~\ref{sec:related}) and eventually conclude the paper and outline future work (Section~\ref{sec:conclusion}).

\section{Method}\label{sec:method}
We generate an undirected multi-graph where nodes are \emph{entities} mentioned in the reference corpus and two nodes are connected by one or more edges if the two entities are mentioned contextually. \emph{Entity mentions} can be identified by entity linking tools (we employed BLINK~\cite{wu2020scalable}), which associate text spans referring to entities with corresponding entries in a knowledge base (in this work Wikipedia).
To define when two entities are mentioned contextually, we introduce the concept of \emph{frame}. In linguistics, a \emph{semantic frame}~\cite{fillmore2001frame} represents a set of concepts in a sentence which are related to a same action, event, or situation, usually described by a verb. In this paper, we generalize the concept of frame considering the possibility to be evoked by any text portion delimited by some rules. Therefore a \emph{frame} can be evoked by a \emph{sentence} or even a \emph{paragraph} or a \emph{document}, depending on the level of granularity we want to explore. We associate each edge of the graph to the location (document, sentence, text span) of the corresponding frame in the reference corpus.

Formally, we denote the multi-graph with $G=(V,E)$ where $V$ is the set of vertices, composed by all entities mentioned in the reference corpus, and $E$ is the set of edges, where an edge is a triple $e=(u,v,f)$ where $u,v\in V$ represent co-occurrent entities and $f$ is the reference to the frame where they occur. We also define $F(u,v)$ as the set of frames that contain mentions $u$ and $v$ in the reference corpus. 

Given a claim $c$, we first extract all entity mentions from $c$ by means of an entity linking tool~\cite{wu2020scalable}. We refer to the set of entity mentions in $c$ as $M_c$. The evidence can be traced by extracting from $G$ all paths $p$ that satisfy the following conditions:

\begin{itemize}
    \item $p$ begins and ends with entities in $M_c$;
    \item each entity in $M_c$ appears in $p$ exactly once;
    \item the distance in $p$ (number of hops) between two consecutive entities in $M_c$ must not be above a predefined threshold $l$ 
\end{itemize}

Each path outlines the set of frames (e.g., evoked by sentences) that might represent evidence for the claim. Specifically, given a path, the text associated to edges of the path form the candidate evidence for or against the claim. As a special case, when the claim contains just two entities $u$ and $v$, all paths of length not above $l$ from $u$ to $v$ (or, equivalently, from $v$ to $u$) are extracted.

\begin{figure}
\includegraphics[width=\textwidth]{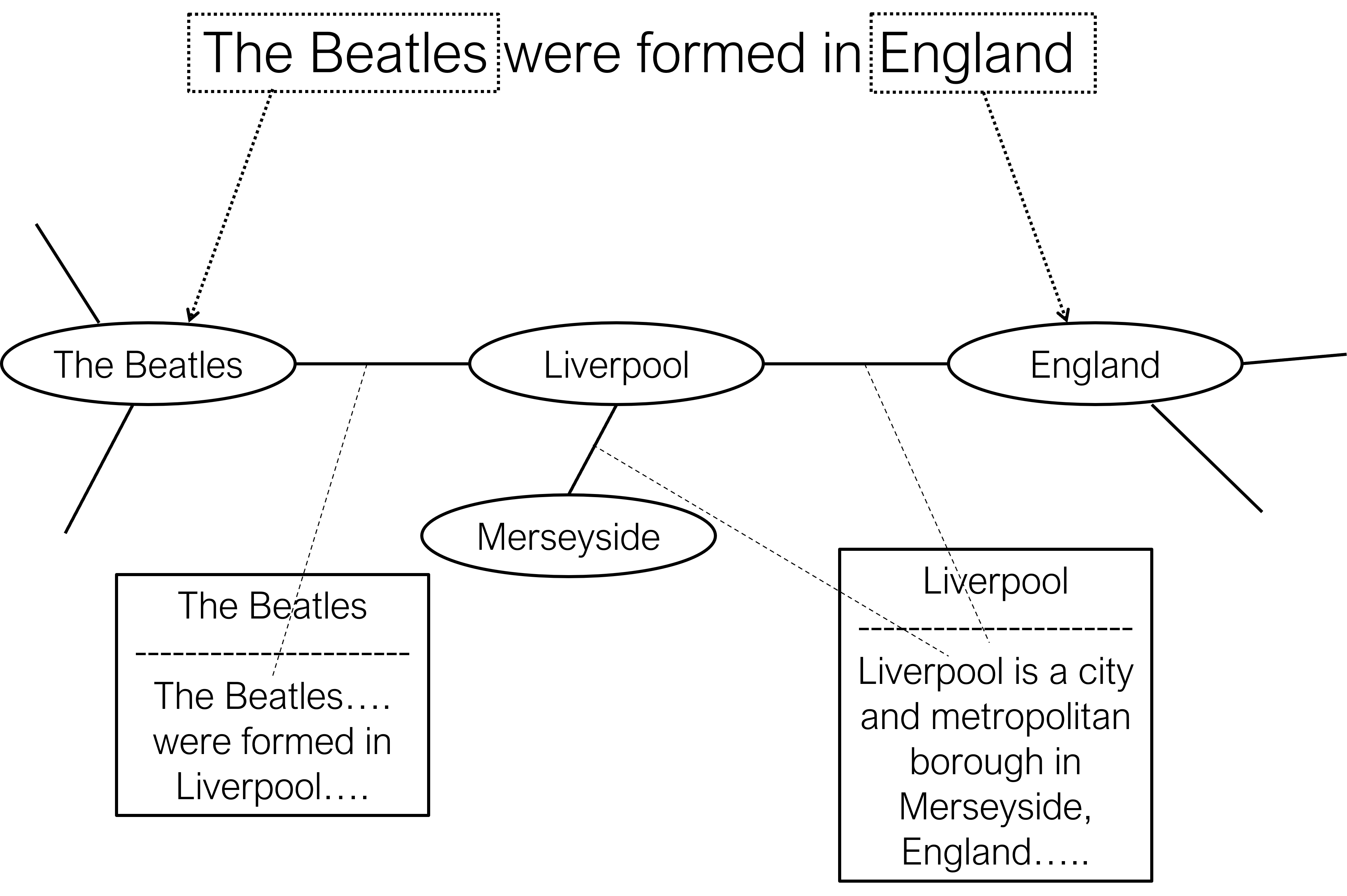}
\caption{An example of claim whose evidence is outlined by a path in the graph.}\label{fig:method}
\end{figure}

Fig.~\ref{fig:method} gives a general idea of the proposed method. In the center we show a fragment of the graph, where nodes are entities, and edges connect entities that are mentioned contextually. Edges are associated to the sentences where the entities co-occur in the reference corpus (pages in the bottom). The sentence in the top is the claim. It contains two named entities (``The Beatles'' and ``England'') that are disambiguated and connected to nodes in the graph. The path that connects the two entities outlines the evidence for the claim, which is represented by the sentences associated to its edges.

The problem of finding paths that connect a set of nodes generalizes the Hamiltonian path problem, which is NP-hard ~\cite{hartmanis1982computers}.
A Na\"ive algorithm that enumerates all paths containing nodes in $M_c$ would have exponential complexity. Moreover paths alone might not cover all cases. In some cases interconnecting entities through a tree might be more appropriate. For these reasons we decided to renounce to enumeration and focus on spotting the part of the graph that contains valid paths.

We limit $l$ to be equal to $2$ and consider three types of nodes: \emph{mentioned nodes}, i.e. nodes of $G$ that represent entities in $M_c$, \emph{between nodes}, i.e. nodes that are not entities in $M_c$ but are between two entities in $M_c$ and all the remaining nodes, namely \emph{unrelated nodes}. We remove unrelated nodes and all incident edges, and consider the remaining subgraph $G_c=(V_c,E_c)$ of $G$. We collect $\bigcup_{(u,v) \in E_c}{F(u,v)}$ as the candidate evidence for claim $c$. We add to this set all sentences of pages that correspond to entities mentioned in the claim.

\section{Experimental analysis}\label{sec:experiments}

We performed an experimental analysis aimed at assessing the effectiveness of the graph exploration in retrieving documents related to a claim. We considered a subset of claims from the FEVER dataset~\cite{thorne_fever_2018}. FEVER contains 185K claims manually annotated with the information concerning whether they are supported or refuted (or neither of them) by a reference corpus made by 5.4M Wikipedia pages. For each supported or refuted claim FEVER provides all lines of evidence, consisting in all possible set of sentences that support or refute the claim. Since our work is specifically focused on finding cross-document evidence, we selected claims whose evidence is across different pages. We also discarded claims that have less than two disambiguated entities and claims that contain entities that are too general (more than 1000 mentions in the corpus).  

We implemented the proposed tool in Python 3.7. We employed the package \emph{spacy} for sententization and tockenization and the BLINK implementation from the authors\footnote{https://github.com/facebookresearch/BLINK} for entity linking. We considered sentences as frames. To manage the size of the graph (5.4M nodes and 68.8M edges), we stored it page by page in a key-value berkeleydb (version 5.3) database, where the key is the page ID and the value is the graph associated to the page. We also built an inverted index that associates entity mentions with the pages and the sentences they occur in. We applied the following procedure for each claim $c$: compute the set $M_c$ of entities mentioned in $c$; retrieve from berkeleydb all graphs of pages that mention entities in $M_c$; get rid of edges of non-relevant sentences and combine the graphs; apply our tool described in Section~\ref{sec:method}.

We evaluated the proposed method, namely GraphRetrieve, against two baselines. The first, namely EntityRetrieve, returns all sentences of documents (Wikipedia pages) corresponding to disambiguated entities in the claim. The second, MentionRetrieve, return all sentences that mention at least one entity in the claim. We also combined the two baselines (Entity+MentionRetrieve) by merging sentences from both Methods.

Results are reported in Table~\ref{tab:results}. We computed the ``hit rate'' as the percent of retrieval successes, i.e. the percent of claims for which all sentences of at least one line of evidence have been retrieved. This value has to be balanced with the amount of data retrieved, since a higher volume increases the chances to make a hit but would penalize the classification task. EntityRetrieve returns a small number of sentences (27.7 on the average, distributed across an average of 2.1 documents) but achieves the lowest hit rate (39.7\%). Entity+MentionRetrieve achieves the highest hit rate (78.7\%) but returns a significant amount of data (333.3 sentences over 250.5 documents).  GraphRetrieve achieves a fairly similar hit rate (70.1\%) with a moderate volume of data (128.5 sentences), which is a little more than one third of the one retrieved by Entity+MentionRetrieve.

\begin{table}
\begin{center}
\caption{GraphRetrieve achieves a hit rate slightly lower than Entity+MentionRetrieve, with a little more than one third of retrieved sentences.}\label{tab:results}
\begin{tabular}{|l|c|c|c|}
\hline
Method            & \multicolumn{1}{l|}{avg. sentences} & \multicolumn{1}{l|}{avg. documents} & \multicolumn{1}{l|}{hit rate} \\ \hline \hline
EntityRetrieve    & \textbf{27.7}                                & \textbf{2.1}                                 & 39.7\%                          \\ \hline
MentionRetrieve   & 312.4                               & 250.4                               & 47.0\%                          \\ \hline
Entity+MentionRetrieve   & 333.3                               & 250.5                               & \textbf{78.7\%}                          \\ \hline
GraphRetrieve (ours) & 128.5                               & 89.6                                & 70.2\%                          \\ \hline
\end{tabular}
\end{center}
\end{table}

Note that GraphRetrieve completely ignores the semantics of relations between entities. Despite its simplicity, it makes a fairly good job in identifying the correct evidence. It reduces considerably the volume of data retrieved, with a small cost in terms of relevant evidence loss. 

\section{Related work}\label{sec:related}
Claim verification is a fundamental step in fact-checking. Interested readers can find an extensive survey on the whole topic in~\cite{guo2021survey}. Recent claim verification methods can rely on large annotated datasets to train machine learning models and achieve considerable results. Thorne et al.~\cite{thorne_fever_2018} provided FEVER, the first large-scale dataset for claim verification over a reference corpus consisting of $185,445$ claims classified as \emph{supported}, \emph{refuted} or \emph{not enough info} and associated to evidence from a corpus of $5.4$ million Wikipedia pages. They described a pipeline that comprises the information retrieval and the textual entailment components. Recently proposed claim verification systems are mainly based on such a framework~\cite{zhou_gear_2019,zhong_reasoning_2020,yoneda_ucl_2018,soleimani_bert_2020,nie_combining_2019,ma_sentence-level_2019,liu_fine-grained_2021-1,hanselowski_ukp-athene_2018}, where a retrieval component  extracts sentences related to the claim from the corpus (the evidence), and a textual entailment component classifies the claim, based on the retrieved evidence. The retrieval component is usually decomposed in two sub-components: document retrieval, which identifies related documents, and sentence selection, which extracts salient sentences from the retrieved documents. The large size of FEVER enables training machine learning models for the task and obtaining performances that overcome $70\%$ accuracy overall.

The document retrieval step is often shared among different works. A commonly used technique consists in retrieving a set of documents by keyword matching with the document titles~\cite{nie_combining_2019,zhong_reasoning_2020} or calling the MediaWikiAPI\footnote{\url{https://www.mediawiki.org/wiki/API}} 
of noun phrases from the claim~\cite{zhou_gear_2019,soleimani_bert_2020,liu_fine-grained_2021-1,hanselowski_ukp-athene_2018}. Some methods also filter retrieved documents by a classifier based on NSMN~\cite{nie_combining_2019,zhong_reasoning_2020,ma_sentence-level_2019}, a variant of ESIM~\cite{chen_enhanced_2017}, a deep learning architecture based on two bidirectional LSTM (Long Short Term Memory) architectures. The claim and its noun phrases are compared with titles of previously retrieved documents to decide its relevance and filter out irrelevant documents.
The sequence retrieval step is usually performed by a classifier that decides for every sentence of the retrieved documents whether it is related or not to the claim. Some systems employ ESIM~\cite{hanselowski_ukp-athene_2018,zhou_gear_2019}, NSMN~\cite{nie_combining_2019}, or logistic regression~\cite{yoneda_ucl_2018} for this step. More recent systems employ transformers, the last generation language models, such as BERT~\cite{soleimani_bert_2020,liu_fine-grained_2021-1} and XLNet~\cite{zhong_reasoning_2020}.

A limit of the described approaches concerns the document retrieval phase. It gives no guaranty that available evidence for or against a claim is retrieved, since such evidence might be contained in documents whose titles might be loosely related or even not related at all to the claim. More advanced retrieval approaches~\cite{chang_pre-training_2020,samarinas_latent_2020,lee_latent_2019,lewis_retrieval-augmented_2021,2020arXiv200208909G} focus explicitly on the retrieval phase, aiming at selecting relevant content for diverse NLP tasks, including claim verification and question answering. They employ two encoders for embedding the documents and the query (e.g. a claim) into the same space, and perform a cosine similarity search to retrieve candidate documents. Eventually the search is refined by a cross-encoder classifier that combines each candidate document with the query and decides if it is relevant. The search can be performed at a finer level of granularity by considering short passages in place of complete documents. Although the described retrieval approaches have been proved successful in solving NLP tasks, including claim verification~\cite{lee_latent_2019,lewis_retrieval_2021}, they suffer when the evidence is fragmented across several documents, each of them loosely related to the claim. Our approach aims at overcoming this limit by interconnecting sentences of the reference corpus and providing a method for spotting all fragments of candidate evidence at once. 



\section{Conclusion}\label{sec:conclusion}
We considered the problem of retrieving evidence for claim verification in the case when such evidence is distributed across different documents in a reference corpus. Available methods do not handle this case appropriately since each part of evidence is retrieved independently. We interconnected all sentences in the reference corpus into a large graph and investigated whether the evidence can be identified in it as a subgraph. Despite the simplicity of the method, which do not even consider the semantics of the relation between entities, we are able to considerably reduce the amount of candidate evidence with a small loss of relevant text with respect to baseline approaches. As future work we plan to incorporate the semantics of relations between entities to make a more focused search and further improve the method.

\section*{Acknowledgment}
The authors are partially supported by the Italian ``Ministero dell’Universit\`a e della Ricerca'' under the project ``SI-ROBOTICS: SocIal ROBOTICS for active and healthy ageing'' (PON 676 – Ricerca e Innovazione 2014-2020-G.A. ARS01\_01120).

\bibliographystyle{unsrt}
\bibliography{references}

\end{document}